\def\year{2020}\relax
\documentclass[letterpaper]{article} 
\usepackage{aaai20}  
\usepackage{times}  
\usepackage{helvet} 
\usepackage{courier}  
\usepackage[hyphens]{url}  
\usepackage{graphicx} 
\usepackage{graphicx}  
\usepackage{caption}
\usepackage{subcaption}
\title{Physics Guided Machine Learning Methods for Hydrology}
\author{Ankush Khandelwal\textsuperscript{\rm 1}\thanks{corresponding author: khand035@umn.edu}, Shaoming Xu \textsuperscript{\rm 1}, Xiang Li\textsuperscript{\rm 2}, Xiaowei Jia\textsuperscript{\rm 4}\\ \Large \textbf{Michael Stienbach \textsuperscript{\rm 1}, Christopher Duffy \textsuperscript{\rm 3}, John Nieber \textsuperscript{\rm 2} and Vipin Kumar \textsuperscript{\rm 1}}\\ 
\textsuperscript{\rm 1}Department of Computer Science and Engineering, University of Minnesota\\
\textsuperscript{\rm 2}Department of Bioproducts and Biosystems Engineering, University of Minnesota\\
\textsuperscript{\rm 3}Department of Civil and Environmental Engineering, Pennsylvania State University\\
\textsuperscript{\rm 4}Department of Computer Science, University of Pittsburgh\\
}

\begin{document}

\maketitle

\begin{abstract}
Streamflow prediction is one of the key challenges in the field of hydrology due to the complex interplay between multiple non-linear physical mechanisms behind streamflow generation. While physics based models are rooted in rich understanding of the physical processes, a significant performance gap still remains which can be potentially addressed by leveraging the recent advances in machine learning. The goal of this work is to incorporate our understanding of hydrological processes and constraints into machine learning algorithms to improve the predictive performance.
Traditional ML models for this problem predict streamflow using weather drivers as input. However there are multiple intermediate processes that interact to generate streamflow from weather drivers.
The key idea of the approach is to explicitly model these intermediate processes that connect weather drivers to streamflow using a multi-task learning framework. While our proposed approach requires data about intermediate processes during training, only weather drivers will be needed to predict the streamflow during testing phase. 
We assess the efficacy of the approach on a simulation dataset generated by the SWAT model for a catchment located in the South Branch of the Root River Watershed in southeast Minnesota. While the focus of this paper is on improving the performance given data from a single catchment, methodology presented here is applicable to ML-based approaches that use data from multiple catchments to improve performance of each individual catchment
\end{abstract}

\section{Introduction}
Streamflow prediction is one of the key tasks for effective water resource management. A number of physics based models have been developed by hydrologists to model different aspects of the water cycle using physical equations. A major drawback of these models is that they require extensive effort to calibrate for any given geography of interest \cite{Zheng2012,Shen2012}. Moreover, these models are necessarily approximate representations of the underlying physical phenomena, and thus are limited in their ability to predict the physical quantities (fluxes) of interest. In recent years, deep learning techniques have shown tremendous success in a number of computer vision and natural language processing applications. 
There is a huge interest in exploring the utility of these schemes for addressing the well-known limitations of physical models in various scientific domains including hydrology
\cite{Nearing2020,Shen2018,Boyraz2018,Fan2020,Hu2020,Ni2020,Kratzert2018,Kratzert2019,Yang2020,Feng2019,Hu2018,Fu2020,Shen2017}. In some cases, it has been shown that LSTM based models are able to 
out-perform physics based models calibrated for individual catchments. In nearly all of these works, ML models are trained to  use weather variables and catchment characteristics as input to estimate streamflow. 

However there are multiple intermediate processes that interact to generate streamflow from weather drivers. The key idea we explore in this paper is to explicitly model physical principles of the hdyrological cycle that connect weather drivers to streamflow using a multi-task learning framework.
In this paper, we consider the scenario where data about other intermediate variables such as soil-moisture, snow-pack, evapotranspiration are also available during training (but not during testing/prediction phase). Specifically, our goal is to investigate how these variables can be used together to improve the prediction performance. 

\begin{figure}[t]
\centering
\includegraphics[width=0.9\columnwidth]{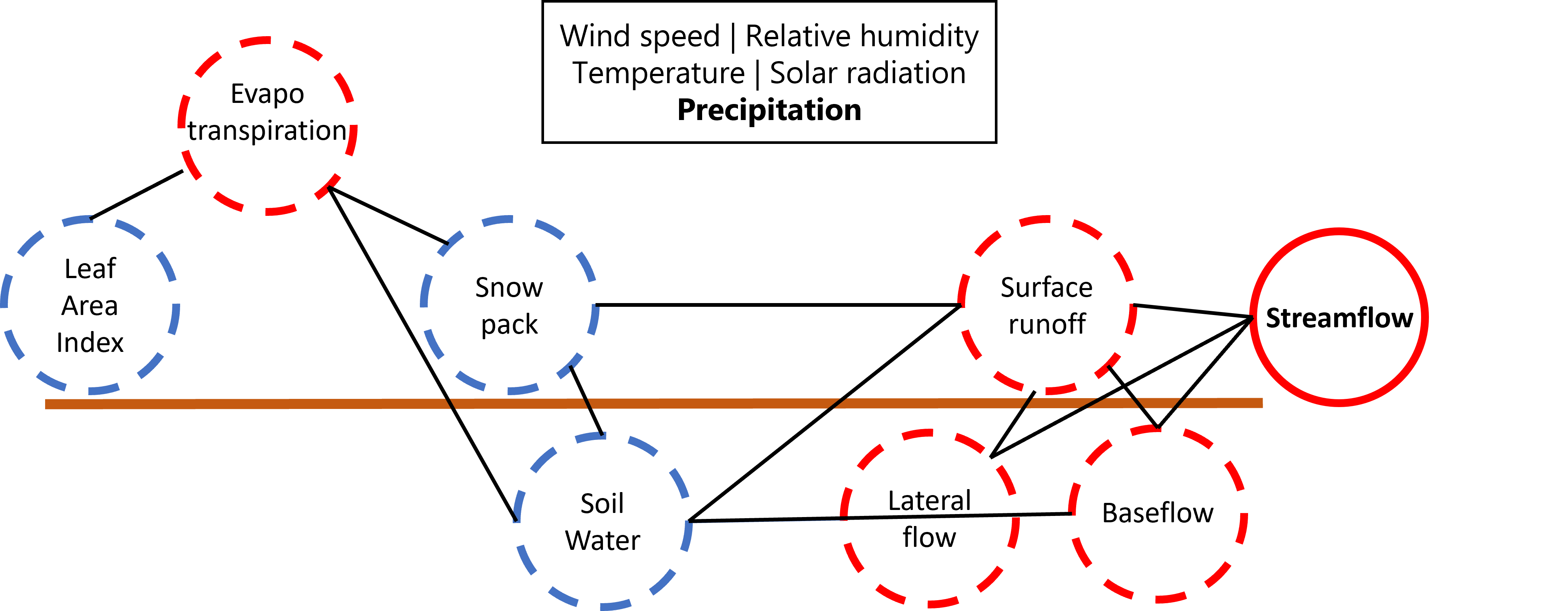} 
\caption{A graphical abstraction of the hydrological cycle. Intermediate variables are represented by dashed circles. Red color denotes state variables where blue color denotes fluxes.}
\label{fig1}
\end{figure}

The hydrological cycle has strong temporal structure and thus time aware deep learning techniques such as RNNs can be used to model different output variables using weather inputs. However, the mapping from weather inputs to variables of interest is very complex, and involves a number of intermediate variables that are inter-connected as shown in Figure \ref{fig1}. Some of the intermediate variables acts as a state of the hydrological system (shown in blue in Figure \ref{fig1}) and govern the response of other variables (also called as fluxes, and shown in red in Figure \ref{fig1}). The relationship between states and fluxes can be seen as hierarchy where states play a significant role in the response of flux variables to weather inputs. For example, for a given amount of rainfall, the amount of surface run-off will depend on how much water is already present in the soil. In other words, if the soil is very wet before rainfall occurs, it will lead to more surface run-off compared to the scenario when the soil is dry. Similarly, for any given temperature distribution over the day, the amount of water available through snow melt will depend on how much snowpack is already present. 

Hence, a framework that captures these relationships has the potential to perform better than directly mapping weather inputs to streamflow. In this paper, we show the utility of explicitly modeling intermediate states as well as fluxes to incorporate the physical relationships between different hydrological processes. The efficacy of the proposed framework is evaluated using a 1000 year simulation dataset created from SWAT model. Our preliminary analysis shows the promise of the proposed approach and provide insights for future directions.

\begin{figure}[t]
\centering
\includegraphics[width=0.9\columnwidth]{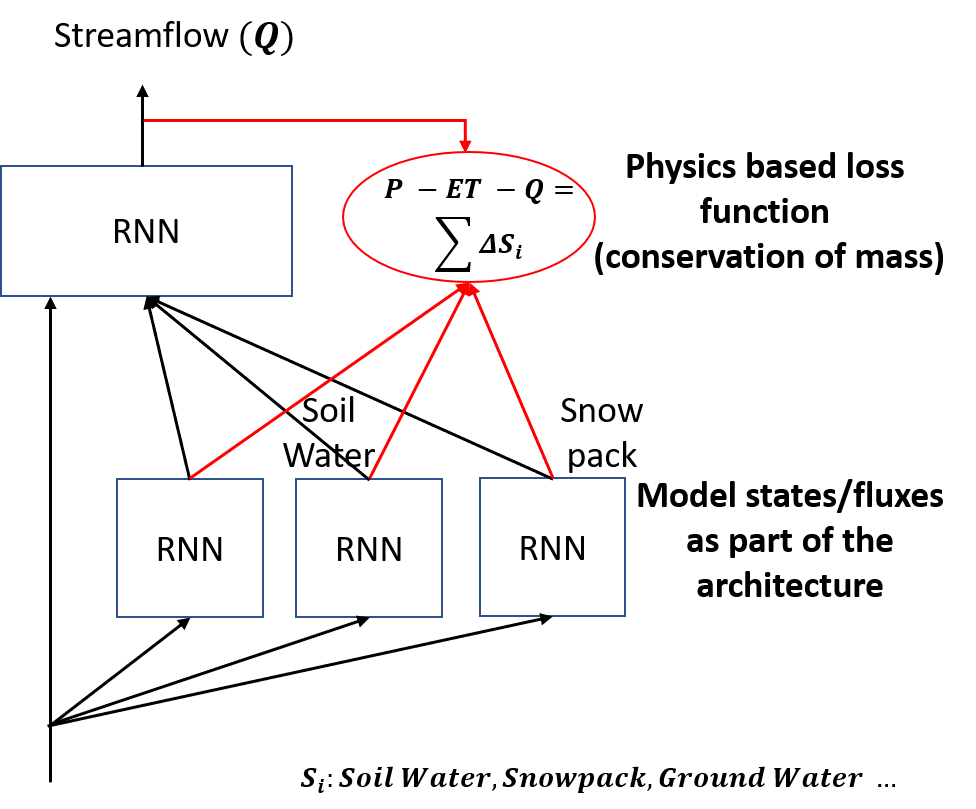} 
\caption{A physics-guided deep learning framework for estimating streamflow.}
\label{framework}
\end{figure}



\begin{figure}[t]
     \centering
     \begin{subfigure}[b]{0.4\columnwidth}
         \centering
         \includegraphics[width=\textwidth]{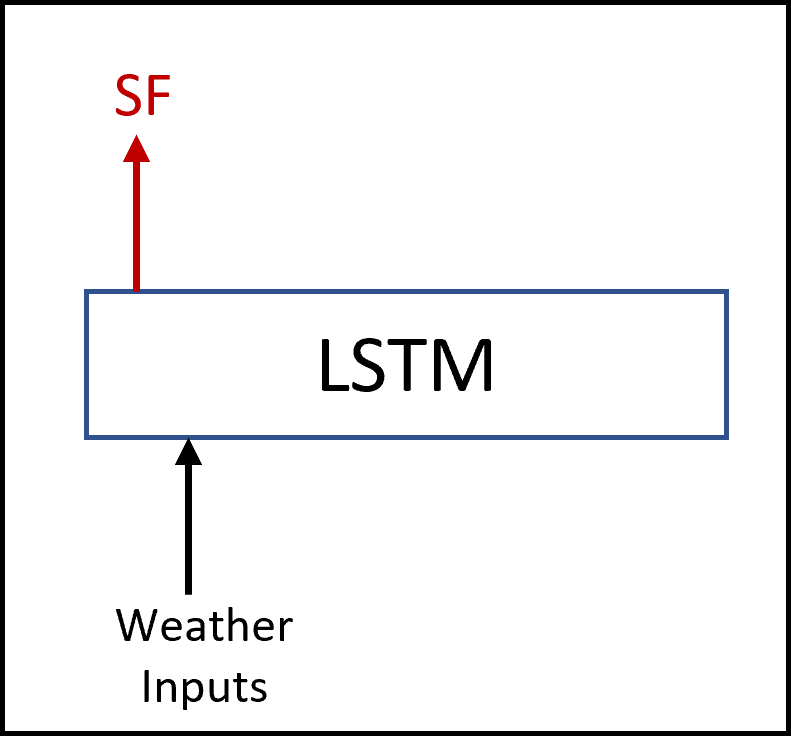}
         \caption{Single Task Learning}
         \label{fig:y equals x}
     \end{subfigure}
     \hfill
     \begin{subfigure}[b]{0.4\columnwidth}
         \centering
         \includegraphics[width=\textwidth]{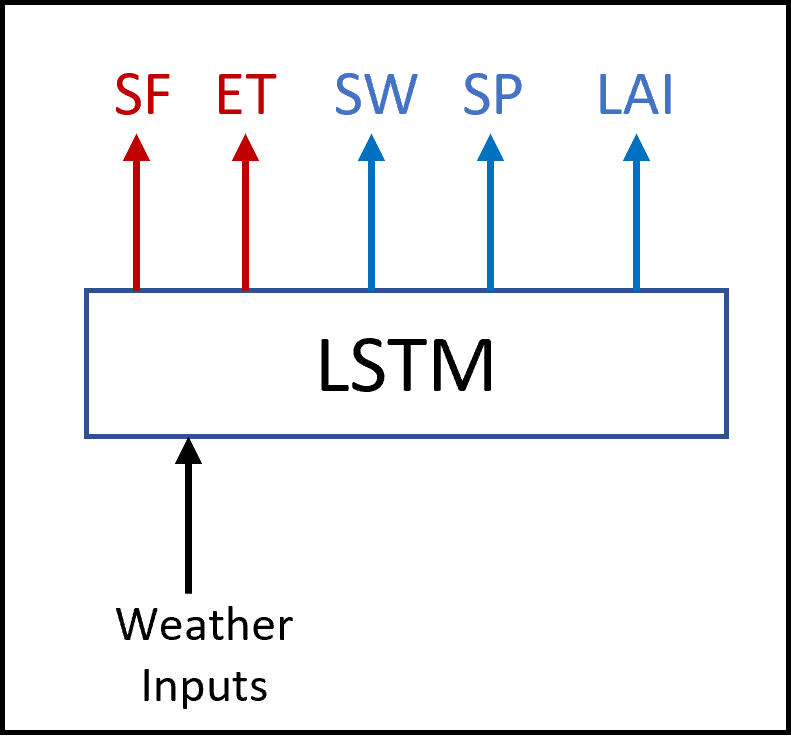}
         \caption{Multi-task Learning}
         \label{fig:three sin x}
     \end{subfigure}
     \hfill
     \begin{subfigure}[b]{0.4\columnwidth}
         \centering
         \includegraphics[width=\textwidth]{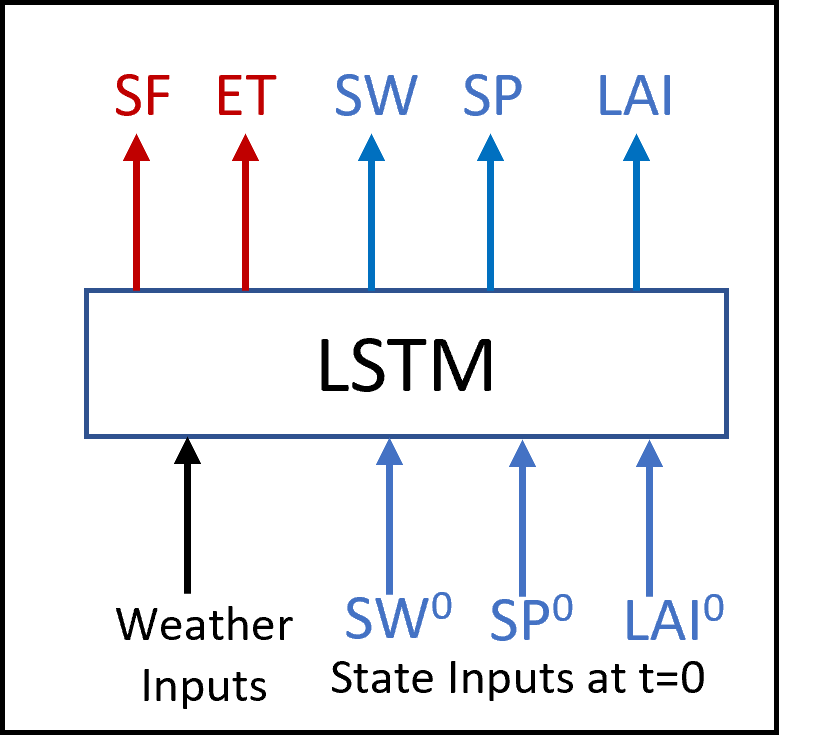}
         \caption{State-Aware Multi-task Learning}
         \label{fig:five over x}
     \end{subfigure}
     \hfill
     \begin{subfigure}[b]{0.44\columnwidth}
         \centering
         \includegraphics[width=\textwidth]{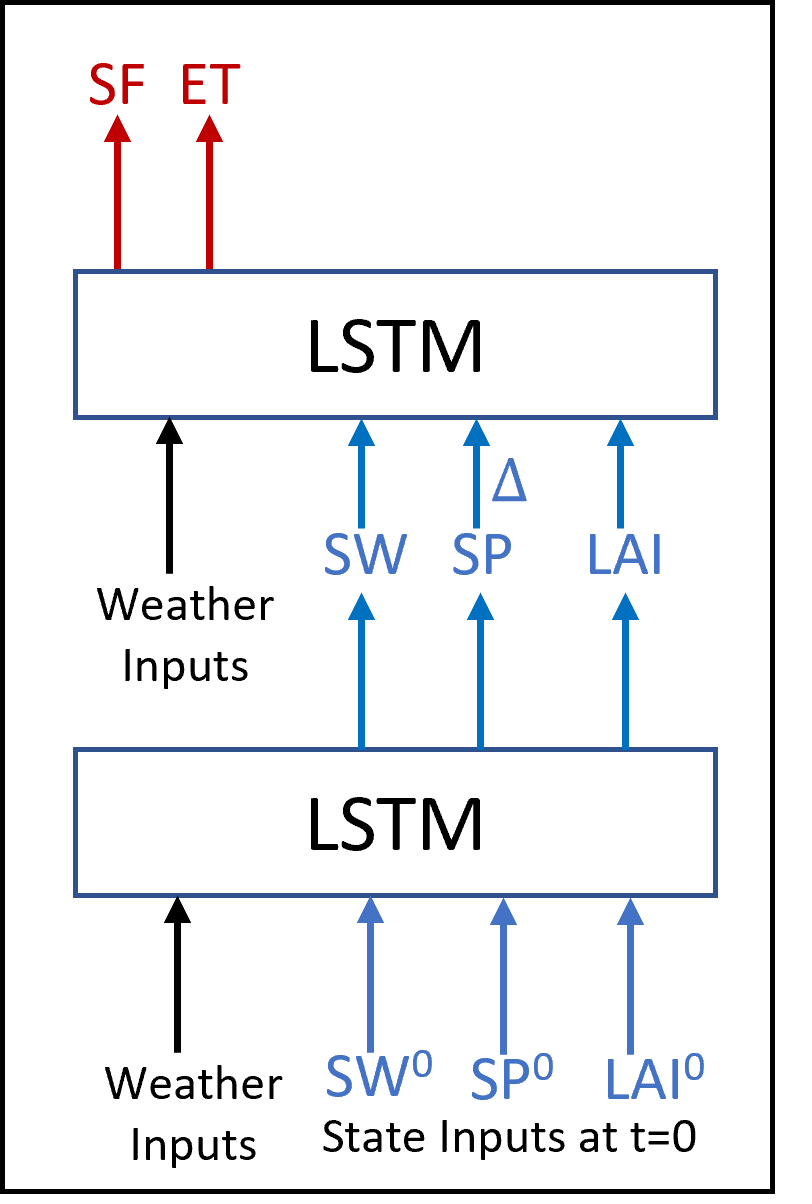}
         \caption{Hierarchical State-Aware Multi-task Learning }
         \label{fig:five over x}
     \end{subfigure}
        \caption{Physics guided deep learning architecture for estimating streamflow. Red color denotes state variables where blue color denotes fluxes.}
\label{fig2}

\end{figure}

\section{Methodology}
Given a timeseries of weather inputs, our goal is to predict streamflow for each timestep. The most intuitive architecture would be an LSTM (or other RNN variants)  to directly map weather inputs to streamflow as shown in Figure \ref{fig2} (a). In this paper, our aim is to explore the impact of using other intermediate state and flux variables to improve modeling of streamflow. Furthermore, we make an assumption that these intermediate variables are not available during prediction phase because observing these variables is costly and time-consuming in real-world scenarios. Hence, our goal is to build models where these intermediate variables are not needed as inputs during the prediction phase. Specifically, we analyze 3 different architecture as shown in Figure \ref{fig2}(b-d) that capture physical concepts of different complexity. In these architectures, the variables depicted in blue represents variables that are used only during the training phase. In other words, during prediction phase, these models only rely on weather inputs and their own predictions. Next, we describe these architecture in detail - 
\newline

\textbf{Multi-task Learning (MTL):} Instead of just predicting streamflow, we propose to train the model to learn other intermediate variables as well. Our hypothesis is that by training related variables together, the model will have better performance and generalization ability in predicting streamflow. Figure \ref{fig2}(b) shows the multi-task learning architecture considered in this paper. 

Since, output variables have different distributions and data uncertainty, a trivial merging of loss terms would lead to sub-optimal training performance. In order to effectively optimize multiple output variables, we use self-paced learning strategy \cite{kraft2020hybrid}. In self-paced learning, an uncertainty term is learned for each task to capture different uncertainties of target variables, which are then used to weigh the loss terms. The self-paced learning strategy is also used for the remaining two formulations discussed below.
\newline

\begin{figure}[ht]
\centering
\includegraphics[width=0.9\columnwidth]{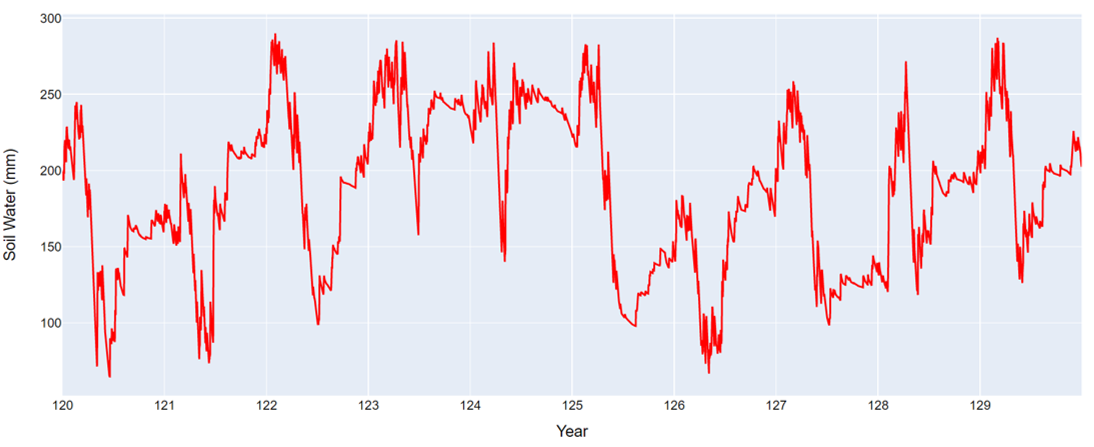} 
\caption{A simulation timeseries of Soil water timeseries from}
\label{SWProfile}
\end{figure}

\begin{figure}[ht]
\centering
\includegraphics[width=0.9\columnwidth]{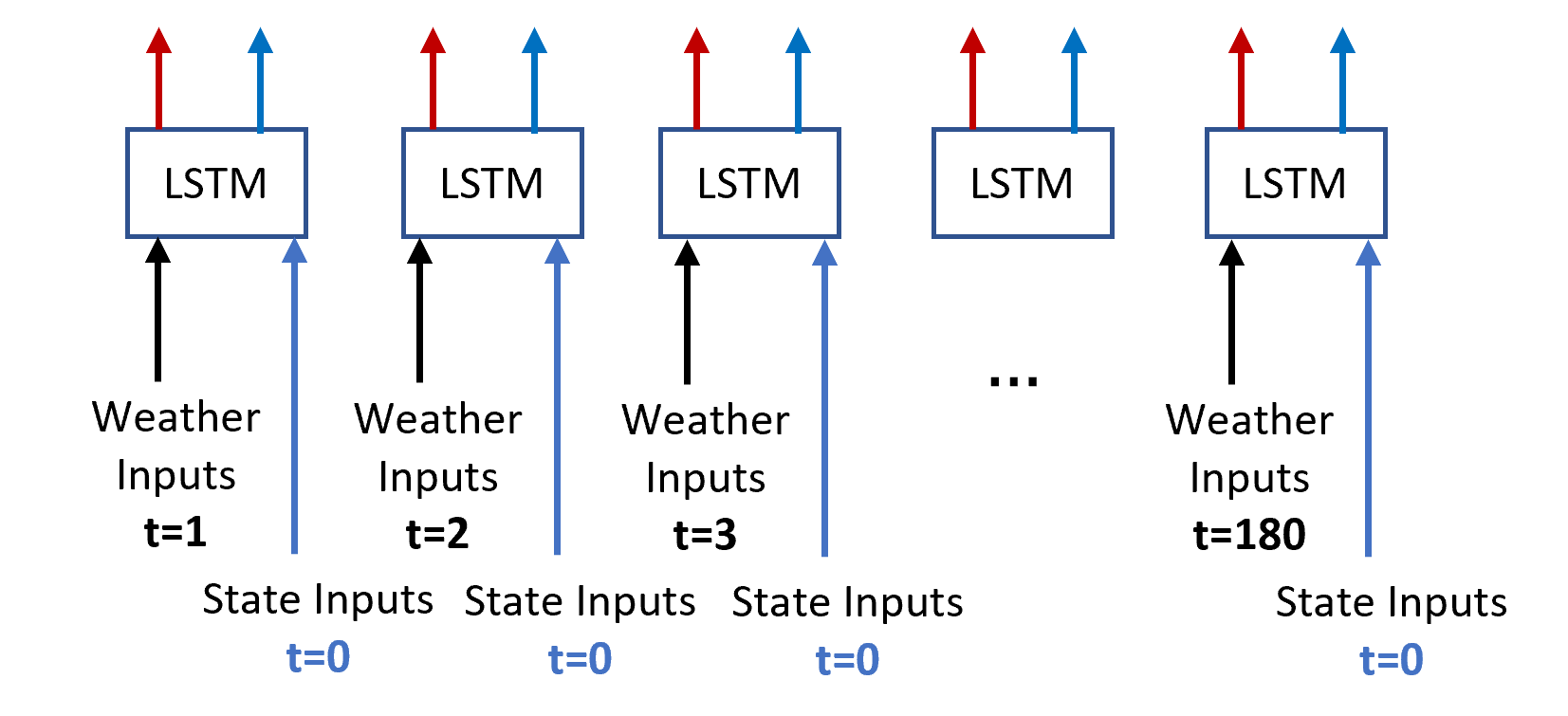} 
\caption{An illustration to depict how initial values are passed while training and prediction.}
\label{training-strategy}
\end{figure}

\textbf{State-Aware Multi-task Learning (SA-MTL):} While the previous architecture considers intermediate variables, it does not differentiate between state and flux variables. State variables have very long memory and can be challenging to model. For example, Figure \ref{SWProfile} shows the variation of Soil Water for a period of 10 years in the simulation dataset (described later). 
In this paper, we propose a new variation to train state variables. Specifically, for a given training sample, along with weather inputs, we provide the value of state variables at the starting day of the sample (replicated for other timesteps in the sequence to maintain input dimensions). This strategy mimics the model initialization step in physical models where the initial conditions determine the output generated by physical models. From a machine learning perspective, this initial constant value avoids the cold start of hidden and cell states and thus improve the temporal modeling of state variables with very long memory. Figure \ref{training-strategy} illustrates this idea for a training sample. 

During prediction phase, we use the predicted values from the previous sample to act as the initial value for the next sample. This is important because our goal is to build models that replicate the physical models but can be easily fine-tuned using ground observations in a real-world setting, where intermediate variables are not regularly monitored and hence cannot assumed to be available. Since, predicted value of the last timestep from the the previous sample is used as initial value for the new sample, we need to assign a default value for the first sample to start the prediction process. In order to address this issue, we add a few samples from the training at the beginning of our prediction set to warm up up the model. Furthermore, we considered assigning a zero value and a random value to begin the prediction phase, and observed that the impact of this initial value dissipates after a few samples, and hence addition of few training samples address the cold start problem.

This initial value concept is also used for the hierarchical variation described below.
\newline



\textbf{Hierarchical State-Aware Multi-task Learning (H-SA-MTL):} As mentioned previously, the value of state variables impact the value of fluxes for any given set of weather inputs. For example, same precipitation amount will lead to different streamflow values depending on the soil moisture already present in the soil. To capture this dependency, we propose a hierarchical network that separates state and flux variables as shown in Figure \ref{fig2} (d). Furthermore, the outputs from the state learning LSTM are used as inputs for the flux learning LSTM. 

Overall, all the architectures have access to same inputs and outputs but each architecture use them differently to capture physical principles. 

\section{Dataset}
\begin{figure}[ht]
\centering
\includegraphics[width=0.9\columnwidth]{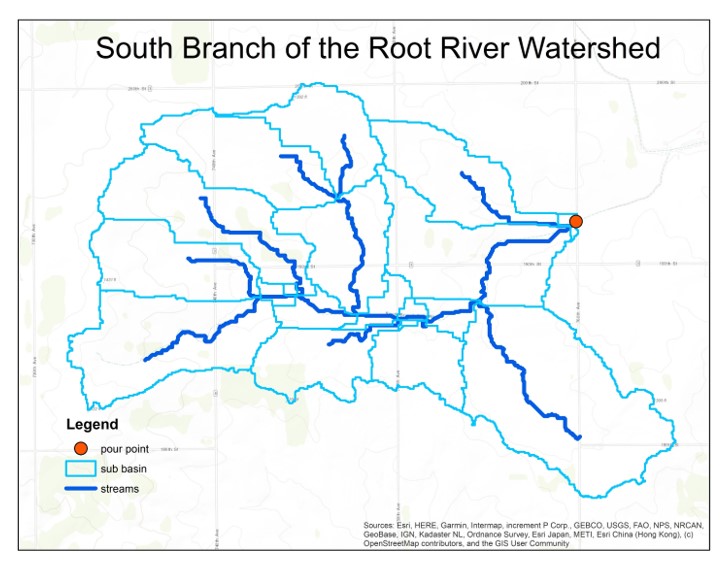} 
\caption{The geographical location of the watershed.}
\label{fig3}
\end{figure}

In this paper, we demonstrate the utility of the proposed architecture using a simulation dataset generated by the SWAT model. Specifically, we created 1000 years of simulation from SWAT which takes 6 weather variables as input (precipitation, minimum day temperature, maximum day temperature, solar radiation, relative humidity, and wind speed) and generates different fluxes and states as output. The model was set up for a watershed in Southwest Minnesota as shown in Figure \ref{fig3}. The weather variables were generated for this region using the weather generator module which is part of the SWAT model. The main goal of the paper is to show the utility of the proposed architectures in emulating the SWAT model. The evaluation of the proposed framework using real-world streamflow data will be pursued in future work.  

\section{Experimental Setup}
To evaluate these architectures, we use first 600 years of the simulation data for training and test its performance on the last 400 years. To generate training samples, we chose 180 days as sequence the length. For training, we set the number of hidden features in LSTM to be 28 and the learning rate was chosen to be 0.001. To evaluate the performance of these architectures, we use RMSE (Root Mean Squared Error) as the evaluation metric. 

\section{Results}
Table \ref{table1} shows the RMSE values for estimating streamflow on testing data by different architectures. 

\begin{center}
\begin{tabular}{ |c|c| } 
 \hline
  Architecture & RMSE \\ \hline
 STL & 0.63 \\ \hline
 MTL & 0.55 \\ \hline
 SA-MTL & 0.40 \\ \hline
 H-SA-MTL & 0.30 \\ \hline
\end{tabular}
\label{table1}
\end{center}

The performance improvement from 0.63 to 0.55 demonstrate that by learning other related variables, the model was able to improve performance in predicting the key target variable, streamflow. The performance continues to improve as architectures include more physical principles. The performance improvement from 0.55 to 0.4 shows the utility of passing initial values of state variables. To further understand the impact of the initial value idea on modeling state variables, we calculate RMSE value for Soil Water using MTL and MTL-SA. The RMSE reduces from 33.06 to 14.25 when initial value is used during training and model predictions are used during testing. Hence, for very long memory state variables, our proposed strategy can significantly improve the performance. Finally, when hierarchy is introduced in the network, the performance further improves from 0.4 to 0.3 which demonstrate the utility of using a separate LSTM for state variables and flux variables. Note that in all these architecture (except the STL baseline), the inputs and outputs are same. Hence, the improvement is coming from architecture choices which highlights the importance of incorporating physical principles. 

\begin{figure}[t]
\centering
\includegraphics[width=0.95\columnwidth]{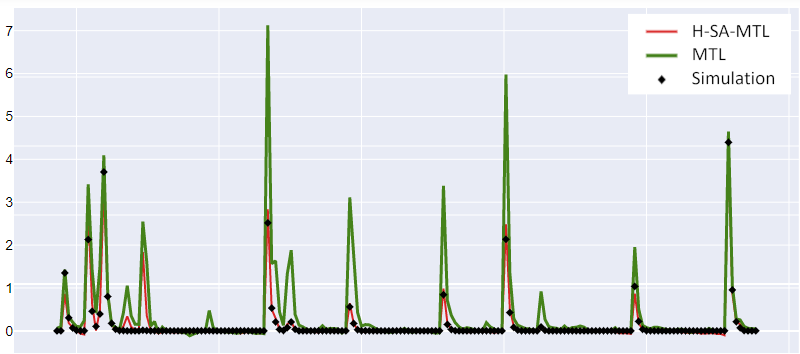} 
\caption{An illustrative example of prediction performance of different architecture configurations. The black diamonds represent streamflow values simulated by SWAT which is being used as reference in our experiments to emulate SWAT. The green circles represent predictions from a traditional multi-task learning architecture (MTL). The red line represents predictions from H-SA-MTL.}
\label{fig4}
\end{figure}

\begin{figure}[t]
\centering
\includegraphics[width=0.95\columnwidth]{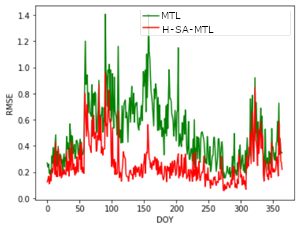} 
\caption{Day specific RMSE value during the testing period for predicting streamflow. Green line represents MTL, and red line represents H-SA-MTL}
\label{fig5}
\end{figure}

Figure \ref{fig4} shows the timeseries of one of the test samples with sequence length of 180 days. As we can see, H-SA-MTL was able to improve the performance on peak values and also reduced the spurious low streamflow values compared to MTL. To provide more insight into the temporal nature of the errors made by different models, we plot RMSE for each day individually as shown in Figure \ref{fig5}. As we can see, the major improvement is coming during summer months due to improved modeling of soil water by our state-aware learning strategy. The lack of improvement during the winter months could be due to insufficient accuracy in modeling snowpack (RMSE in modelling snowpack only improves from 1.08 to 0.98). Hence, further investigation is needed to understand the challenges in modeling snowpack compared to soilwater. 

\section{Summary and Future Work}
In this paper, we presented a physics guided deep learning architecture to improve the performance on the task of predicting streamflow from weather variables. The key idea is to model intermediate states and fluxes of the hydrological cycle explicitly in the model architecture. The hierarchical approach allows different processes that change at different temporal scales to be learned using different LSTMs, and thus improve their modeling. The results on the simulation data demonstrate the utility of the hierarchical multi-task learning approach. A number of possible advancements can further improve the utility of the framework. Specifically, constraints such as mass conservation can be added to increase generalizability of the model. While results on simulation data show promising results, the efficacy of the framework in adapting to observation data (a.k.a. fine-tuning) needs to be investigated. 

Note that our focus has been on improving the predictive performance given data from a single catchment. However, as we mentioned earlier, LSTM based models that leverage data from multiple catchments (conditioned by their physical characteristics) are able to out-perform physics based models calibrated for individual catchments \cite{Kratzert2019,li2022regionalization}. To this end, a possible future direction of the proposed approach is to simulation data of intermediate variables to pre-train these multi-catchment models instead of training them from scratch. This will encode more physical information in the machine learning model, and hence can potentially improve performance especially in limited-training data scenarios. 


\section{ Acknowledgments}
The work is being funded by NSF HDR grant 1934721 and 1934548. Access to computing facilities was provided by Minnesota Supercomputing Institute.

\bibliography{main}
\bibliographystyle{aaai}

\end{document}